\documentclass[10pt,letterpaper,twocolumn]{article} 

\usepackage{ol2_arxiv}
\usepackage[draft]{hyperref}
\usepackage{amsmath}
\usepackage{units}
\usepackage{textcomp}

\begin{document}

\twocolumn[ 

\title{Microdeflectometry -- a novel tool to acquire 3D microtopography\\ with nanometer height resolution}

\author{Gerd H\"ausler,$^{*}$ Claus Richter, Karl-Heinz Leitz, and Markus C. Knauer}

\address{Institute of Optics, Information and Photonics, Max Planck Research Group, \\ University Erlangen-Nuremberg, Staudtstr. 7/B2, 91058 Erlangen, Germany\\
$^*$Corresponding author: haeusler@physik.uni-erlangen.de
}

\begin{abstract}
We introduce "microdeflectometry", a novel technique for measuring the microtopography of specular surfaces. The primary data is the local slope of the surface under test. Measuring the slope instead of the height implies high information efficiency and extreme sensitivity to local shape irregularities. The lateral resolution can be better than 1\textmu m, whereas the resulting height resolution is in the range of 1 nm. Microdeflectometry can be supplemented by methods to expand the depth of field, with the potential to provide quantitative 3D imaging with scanning-electron-microscope-like features. \end{abstract} 

]

\noindent We introduce a microscopic adaptation of deflectometry [\citeonline{Ritter}]. Specifically, we modify the so-called "Phase-Measuring Deflectometry" (PMD)[\citeonline{Haeusler}]. Microdeflectometry provides quantitative slope images with lateral resolution better than 1\textmu m and slope resolution in the range of \unit[1]{mrad}. A surface height variation of 1 nm can be detected within the diffraction-limited resolution cell of the optical system. The method is incoherent (low noise) and provides an angular range as big as the aperture of the microscope.

PMD, as in similar methods [\citeonline{Perard}-\citeonline{Bothe}], is based on the observation of mirror images of remote patterns, using the object under test as a mirror. The mirrored patterns are distorted, depending on the shape of the object. Using sinusoidal fringes and a calibrated system, the local slope of the surface can be calculated by standard phase-shifting techniques, at $\sim 10^6$ object points $(x,y)$, within a few seconds. To get the best possible lateral resolution we focus on the surface of the object under test. Due to the limited depth of field (dof), we cannot focus on the object and on the screen at the same time. However, the measured phase will not be altered by the blurring, since the fringes are sinusoidal [\citeonline{Haeusler}, \citeonline{Knauer2004}].

Measuring the local slope instead of the height of an object has significant advantages. Slope measurement is equivalent to (optical) source encoding for high information efficiency [\citeonline{Wagner2005}]. Slope data are more effective in detecting local details and defects than height data [\citeonline{Knauer2006}]. Macroscopic PMD is a now-established technique for measuring macroscopically specular free-form surfaces, specifically eye glass lenses [\citeonline{3ds}]. Other applications are painted car bodies, car window glasses or precision-machined surfaces.

Our aim is to adapt this measuring principle to a microscopic field of view down to \unit[100]{\textmu m}. This opens up new applications. A few examples are shown in this paper: Cutting tools, microoptical elements, and wafers. With high aperture and high resolution, many surfaces that are matte in a macroscopic setup display specular reflection, and these can be measured with microdeflectometry. With high-aperture imaging steep slopes are accessible (up to $\pm$\,\unit[60]{$\!^\circ$}). 
  
\begin{figure}[ht]
\centerline{\includegraphics[width=8cm]{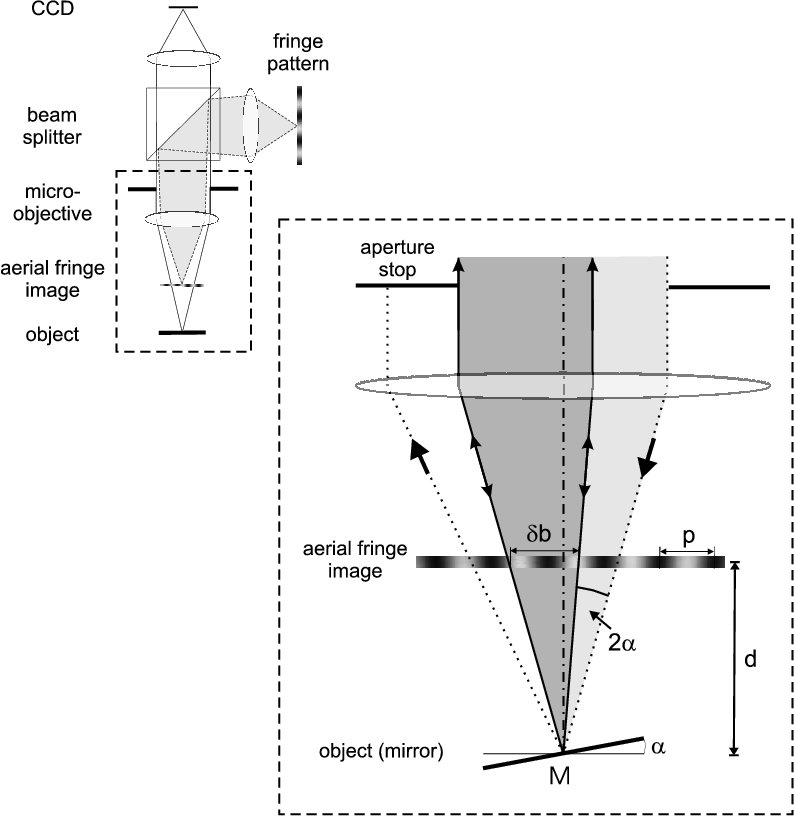}}
\caption{Schematic setup for microscopic PMD.}
\label{muPMD_setup}
\end{figure}  
  
We now explain how to implement microdeflectometry. It is not possible to use a diffusing remote screen, as in macroscopic PMD, to display the fringe pattern, because of the small working distance of a microscope. Our solution is different: As depicted in Fig.\,\ref{muPMD_setup}, we project an aerial image at a distance $d$ from the object. This projection is achieved by means of the same microobjective that acquires both the image of the object and the mirror image of the fringe pattern. We generate this pattern using an electronically controllable light modulator and a beam splitter. This scheme enables an incoherent (low optical noise) and coaxial (no shadowing) illumination with high aperture (high resolution, big angular range) and small working distance.

We briefly explain the imaging scheme for the example of a point $M$ located at the object surface. The cone of imaging rays is represented by the dark gray area. The observed image intensity is given by the intensity within this cone, averaged over the distance $\delta b$. The tilted object causes vignetting and the rays within the light gray cone do not contribute to imaging. The cone of imaging rays depends on the slope $\alpha$. Hence the observed phase $\varphi$ depends on the slope, with $\varphi = 2\pi d \alpha /p$. Note that owing to the vignetting, microdeflectometry is only half as sensitive as macroscopic deflectometry. 

We now discuss the physical limitations. In  [\citeonline{Haeusler}, \citeonline{Knauer2004}] an uncertainty relation as given in Eq.\,(\ref{eq:usr1}) was derived that connects the lateral resolution $\delta x$ and the angular measuring uncertainty $\delta\alpha$, where $\lambda$ is the wavelength and $Q$ is a quality factor, depending solely on the camera noise: 
\begin{equation}
	\delta x \cdot \delta \alpha = \lambda/Q
	\label{eq:usr1}
\end{equation}

Equation\,(\ref{eq:usr1}) appears a little disappointing since we learn that with higher lateral resolution (with higher aperture) the slope uncertainty will become bigger. However, if we put in the experimental value of $Q$\,$\approx$\,1000, we get approximately
\begin{equation}
	\delta x \cdot \delta \alpha \approx 1 \text{nm}
	\label{eq:usr2}
\end{equation}

Equation (\ref{eq:usr2}) reveals a remarkable result: The minimum height variation $\Delta z$ that can be detected within the diffraction-limited resolution distance of the microscope is constant and in the range of only \unit[1]{nm}. (It should be mentioned that there are systems for pointwise macroscopic measurement [\citeonline{ptb}] with $Q > 10^5$ and, hence, with a much smaller $\Delta z$.) 

\begin{figure}[ht]
\centerline{\includegraphics[width=6cm]{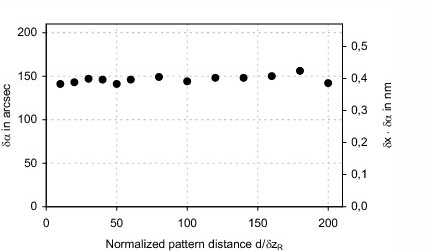}}
\caption{Repeatability, measured on a planar mirror. Displayed are the angular measuring uncertainty $\delta \alpha$ and the corresponding height uncertainty $\delta x \cdot \delta \alpha$ vs. the normalized pattern distance. $\delta z_R$ is the Rayleigh depth of field. Objective: 50x0.85\,.}
\label{davsd}
\end{figure}

One might doubt that Equation (\ref{eq:usr2}) is still valid for high-aperture microdeflectometry, since we choose the distance $d$ bigger than a hundred times the dof. The big distance keeps the relative change of $d$ (caused by height variations of the object) small. Yet the result of our experiments shown in Fig.\,\ref{davsd} demonstrates that the relation is valid even for very high aperture and a very small dof. This is possible because we scale up the fringe period $p$ linearly with the distance $d$. In order to maintain a high contrast in the observed fringe image, the distance $\delta b$ must not exceed $p/2$ [\citeonline{Knauer2004}].  

\begin{figure}[ht]
\begin{center}
	\begin{tabular}{rcl}
	
	(a) & \includegraphics[width=5 cm]{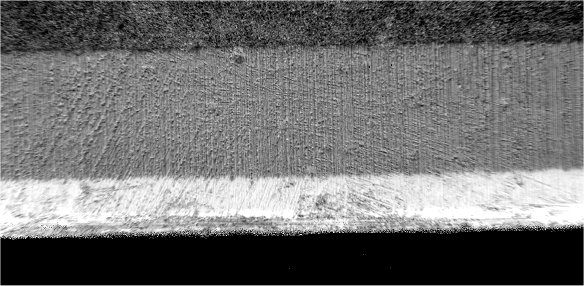} & \includegraphics[width=1.55 cm]{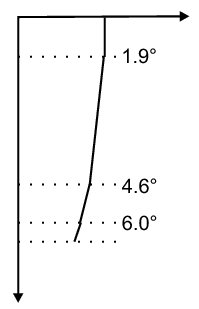} \\
	(b) & \includegraphics[width=5 cm]{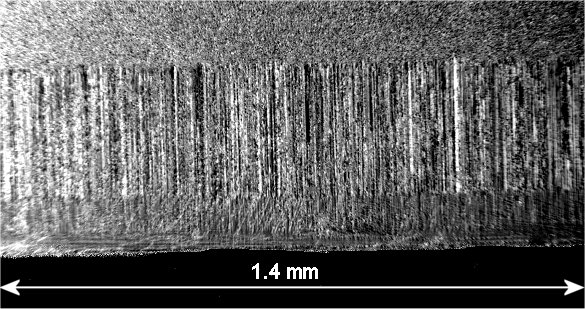} & \includegraphics[width=1.55 cm]{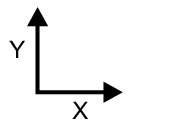}\\

	\end{tabular}
\end{center}
\caption{Razor blade. (a) Local slope in y direction, (b) in x direction. Objective 5x0.12.}
\label{rk_y}
\end{figure}

Now we present some examples of intensity-encoded slope images. The first example is a razor blade (Fig. \ref{rk_y}). The measuring field is \unit[1.4]{mm} (objective 5x0.12). Figure \ref{rk_y}\,(a) displays the different slope angles of the blade, in the \textit{y} direction. The measured angles are close to the reference values of the manufacturer (\unit[2.0]{$\!^\circ$}, \unit[4.5]{$\!^\circ$} and \unit[6.0]{$\!^\circ$}). In Fig. \ref{rk_y}\,(b) the slope in \textit{x} direction is shown, with machining traces emphasized. 

\begin{figure}[ht]
\begin{center}
	\begin{tabular}{rc}
	(a)& \hspace{0.5cm} \includegraphics[width=5 cm]{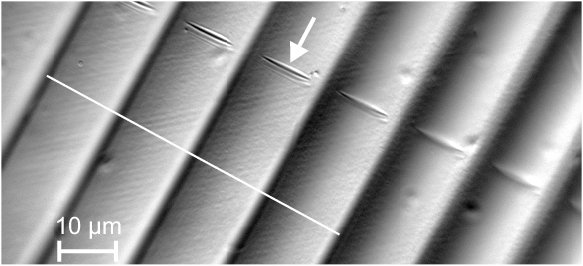} \\
	(b)& \includegraphics[width=6 cm]{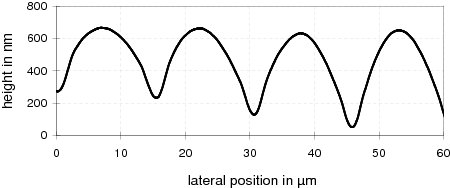} \\
	\end{tabular}
\end{center}	
\caption{(a) Slope image of a microlens array; (b) cross section through height map (by integration).}
\label{ml}
\end{figure} 

The second example is an array of cylindrical micro-lenses (Fig. \ref{ml}\,a). The measuring field is \unit[105]{\textmu m} (objective 50x0.85). The edges of the lenses have a slope of $\sim$ \unit[11]{$\!^\circ$}. The scratches (see arrow) are $\sim$ \unit[40]{nm} deep. In Fig. \ref{ml}\,(b) a cross section of the height map obtained by (2D) integration of the slope data  [\citeonline{Lowitzsch2005}] is depicted. The profile data are degraded (error \textless 10\%) by the still-preliminary calibration of the system. On the right side of the profile there is a systematic error due to defocusing. 
This problem will become obsolete with the solution of the dof problem that we will address now.

\begin{figure}[t]
\centerline{\includegraphics[width=8.1 cm]{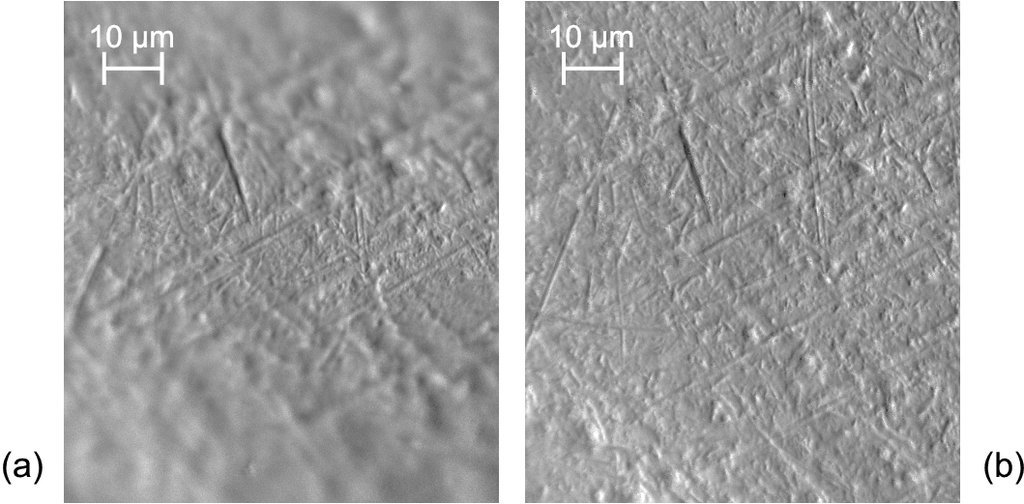}}
\caption{Microdeflectometry with additional expansion of depth of field. Slope image (x direction) of a used bearing ball. Height range approx. 17 \textmu m (30-fold dof, objective 50x0.85). To reduce the dynamical range of the picture, only the deviation of the measured slope from the slope of the sphere is displayed. (a) standard measurement; (b) measurement with about 30-fold expanded dof.}
\label{ste}
\end{figure}
Figure\,\ref{ste} demonstrates that it is possible to combine microdeflectometry with the expansion of the depth of field. The result is an scanning electron microscope (SEM)-like image, which additionally provides quantitative 3D data. The method is based on "image depuzzling", as described in [\citeonline{HaeuslerDOF}]. We extract and combine slope data from sharp image areas acquired at different focus positions.

\begin{figure}[t]
\centerline{\includegraphics[width=6 cm]{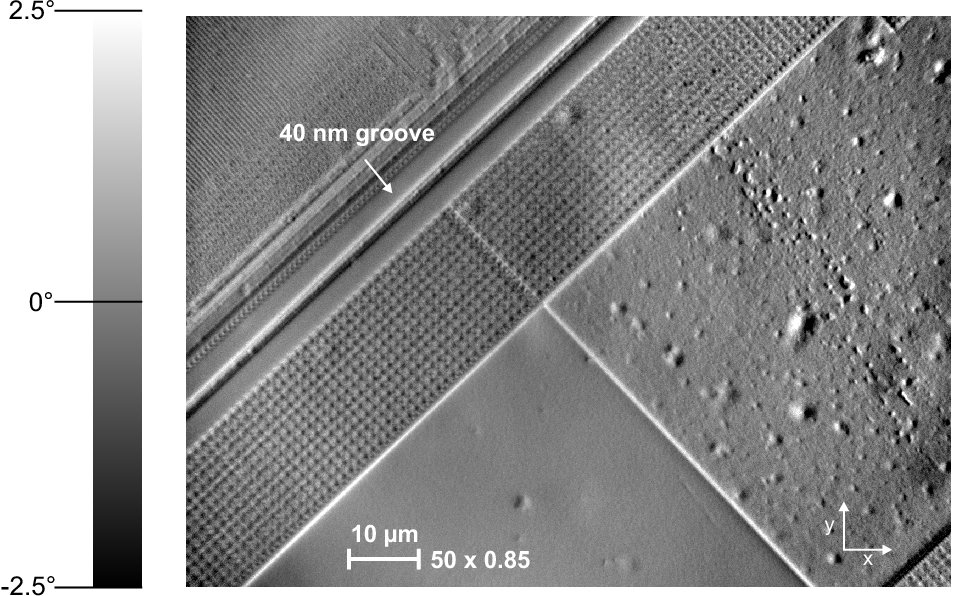}}
\caption{Wafer section. Local slope in y direction.}
\label{wafersy}
\end{figure}

We conclude that microdeflectometry displays several interesting features:

(a) The inherently measured signal is the slope. Slope encoding is what makes differential interference contrast [\citeonline{DIC}]) and SEM images so intriguing. 
The same is true for microdeflectometry images, as (hopefully) shown in Fig.\,\ref{wafersy}.
The slope encoding enhances extremely small depth variations of the surface, down to the nanometer and even sub-nanometer regime. From the slope, the shape $z(x,y)$ can be calculated with extremely low noise [\citeonline{Lowitzsch2005}], since integration is a noise-suppressing operation. (We admit that steps at an object surface cannot be measured. Most smooth surfaces do not display steps, fortunately). High-accuracy quantitative microdeflectometry still requires an improved calibration procedure, and investigations on the influence of diffraction.

(b) The noise level is small. Microdeflectometry can be implemented with high illumination aperture, so residual coherent noise artifacts will have very low contrast. Moreover, the high aperture enables a high dynamical range of the slope measurement. In our experiments we achieved a dynamical range of up to 2500. This is much better than what can be achieved in intensity images.

(c) The lateral resolution and the angular measuring uncertainty are coupled via an uncertainty relation, so that the corresponding parameters can be adapted as required by the application. For eye glass measurements the accuracy  of the refractive power (curvature) is more valuable than the lateral resolution. For semiconductor applications, the lateral resolution is more important.
 
(d) Depth of field: the small dof is a major drawback of microscopic imaging. It is intriguing and a challenge to create SEM-like images (big dof) \textit{optically} with microdeflectometry. We have demonstrated that this is possible, by combining microdeflectometry with the expansion of the dof. In fact, the expansion of the dof with microdeflectometry is easier than with conventional microscopy. 

Microdeflectometry provides appealing pictures, with high lateral resolution, nanometer sensitivity for local surface features, low noise and quantitative 3D features. The technology is simple, incoherent, and has the potential to compete against interferometry.

\end{document}